\documentclass{article}

\usepackage{PRIMEarxiv}

\usepackage[utf8]{inputenc} % allow utf-8 input
\usepackage[T1]{fontenc}    % use 8-bit T1 fonts
\usepackage{hyperref}       % hyperlinks
\usepackage{url}            % simple URL typesetting
\usepackage{booktabs}       % professional-quality tables
\usepackage{amsfonts}       % blackboard math symbols
\usepackage{nicefrac}       % compact symbols for 1/2, etc.
\usepackage{microtype}      % 
\usepackage{lipsum}
\usepackage{fancyhdr}       % header
\usepackage{graphicx}       % graphics

\usepackage{comment} %%%%%%%%%%%%%%%

\usepackage{tabularx}
\usepackage{rotating}
\usepackage{amsmath}
\usepackage{amssymb}
\usepackage{caption}
\usepackage{subcaption}
\usepackage{svg}
\usepackage{url}
\usepackage{multirow}
\usepackage{makecell}
\usepackage{xcolor}
\usepackage{mathtools}  
\usepackage{amsthm}
\usepackage{bookmark}

\DeclarePairedDelimiter\floor{\lfloor}{\rfloor}

\graphicspath{{media/}}     % organize your images and other figures under media/ folder

\title{Learning in Log-Domain: Subthreshold Analog AI Accelerator Based on Stochastic Gradient Descent
}

\author{
  Momen K Tageldeen, Yacine Belgaid, Vivek Mohan, Zhou Wang, Emmanuel M Drakakis    \\
  Imperial College London\\
  \texttt{momen.tageldeen18@alumni.imperial.ac.uk}, \texttt{e.drakakis@imperial.ac.uk}  \\
}

\begin{document}
\maketitle

\begin{abstract}

The rapid proliferation of AI models, coupled with growing demand for edge deployment, necessitates the development of AI hardware that is both high-performance and energy-efficient. In this paper, we propose a novel analog accelerator architecture designed for AI/ML training workloads using stochastic gradient descent with L2 regularization (SGDr). The architecture leverages log-domain circuits in subthreshold MOS and incorporates volatile memory. We establish a mathematical framework for solving SGDr in the continuous time domain and detail the mapping of SGDr learning equations to log-domain circuits. 
By operating in the analog domain and utilizing weak inversion, the proposed design achieves significant reductions in transistor area and power consumption compared to digital implementations. Experimental results demonstrate that the architecture closely approximates ideal behavior, with a mean square error below 0.87\% and precision as low as 8 bits.  Furthermore, the architecture supports a wide range of hyperparameters.  This work paves the way for energy-efficient analog AI hardware with on-chip training capabilities.

\end{abstract}


\section{Introduction}

In recent years, artificial intelligence (AI) has become an integral part of daily life, serving as a transformative tool across various professional domains \cite{Gao2019} and driving personal applications through advancements in transformer models that power large language models (LLMs) \cite{NIPS2017_3f5ee243}. However, both training and inference of AI models demand substantial computational and energy resources, which are becoming increasingly challenging to access \cite{Coates2013, anthony2020carbontracker}. While server-class GPUs are effective for training, their energy inefficiency \cite{Andri2016} and high costs present significant barriers \cite{strubell-etal-2019-energy}. Additionally, the environmental impact of energy-intensive AI systems has raised critical concerns about their role in exacerbating climate change \cite{anthony2020carbontracker}.

Amdahl's law predicts that performance and efficiency gains are best achieved through innovative application-specific accelerator architectures rather than scaling up multi-core general-purpose processors \cite{Xiao2020}. Consequently, application-specific integrated circuits (ASICs), both digital and analog, have emerged as critical solutions for enabling high-efficiency training and inference of artificial neural networks \cite{Xiao2020,Sze2017,Reuther2020}. 

Digital accelerators are widely adopted for training workloads. Notable examples include the Brainwave Neural Processing Unit (NPU) \cite{Chung2018}, Google’s Tensor Processing Unit (TPU) \cite{Jouppi2017}, and low-precision inference accelerators such as YodaNN \cite{Andri2016}, the Unified Neural Processing Unit (UNPU) \cite{Lee2018}, and BRein Memory \cite{Ando2018}. These accelerators leverage specialized architectures to optimize the execution of artificial neural networks, achieving significant performance gains by reducing computational overheads and improving throughput. For instance, the first-generation TPU achieves a peak throughput of 92 tera-operations per second (TOPS) with a thermal design power of 28–40 watts \cite{Jouppi2017}. 

However, despite their energy efficiency at the architectural level, digital accelerators still consume considerable amounts of power due to their reliance on high-speed clocking, frequent memory access, and large-scale digital circuit implementations. For example, training a large language model like GPT-3, which has 175 billion parameters, has been estimated to consume approximately 1,287 megawatt-hours (MWh) of electricity \cite{patterson2021carbonimpact}. These high energy demands raise concerns about their environmental impact and make server-class digital accelerators unsuitable for energy-constrained environments, such as edge devices \cite{strubell-etal-2019-energy, anthony2020carbontracker}.

Analog accelerators, which enable in-memory computation, address the power consumption challenges of digital accelerators by offering significant energy efficiency gains \cite{Xiao2020}. These accelerators, operating in the analog domain, perform operations directly within memory, reducing the need for data movement and associated energy costs. However, many existing analog accelerators are built on emerging memristive devices, such as Resistive RAM (ReRAM) and Phase Change Memory (PCM), which introduce challenges due to their non-idealities and non-linear characteristics. These issues make them difficult to model, characterize, and integrate into robust systems \cite{Xiao2020}. Hybrid approaches, such as the Programmable Ultra-efficient Memristor-based Accelerator (PUMA), have demonstrated superior performance for inference tasks, but they remain limited to inference and are not yet suitable for training workloads \cite{10.1145/3297858.3304049}.

This paper presents a novel analog AI accelerator for training machine learning models, leveraging ultra-low power subthreshold MOS circuits. The proposed accelerator supports stochastic gradient descent with L2 regularization (SGDr) by formulating its computations in continuous time (SGDr-CT). Unlike conventional digital accelerators that require frequent memory access during training, our design uses volatile memory, requiring memory access only at the conclusion of training. Additionally, unlike analog accelerators based on emerging memristive devices, the proposed design utilizes mature CMOS technology. The performance of the accelerator was simulated and validated using the AMS 0.35 µm process, demonstrating its potential for efficient and scalable on-device training.

\section{Background}

\begin{figure}[!ht]
  \centering
  \includegraphics[width= 0.6 \textwidth]{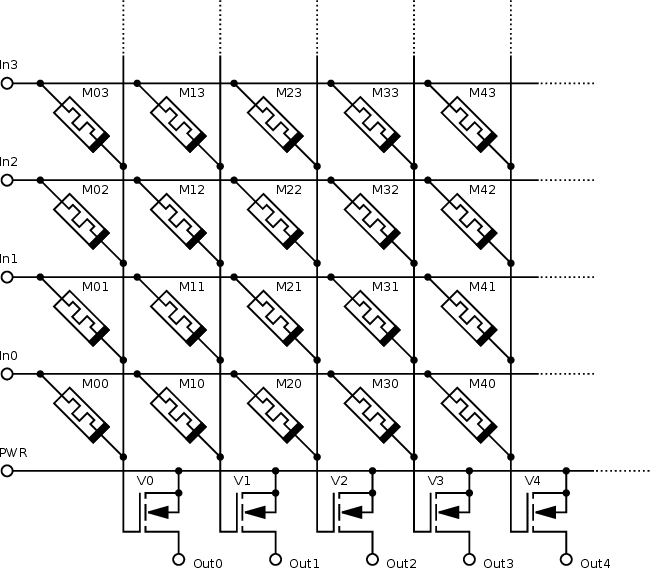}
  \caption[Shows the basic concept of a memristive crossbar.]{The basic concept of a memristive crossbar. The current accumulated at each column represents the output of a single artificial neuron before the activation function. Image: MovGP0 at the German-language Wikipedia, licensed under CC BY-SA 3.0, via Wikimedia Commons\footnotemark.}
  \label{fig:Crossbar}
\end{figure} \footnotetext{source: https://commons.wikimedia.org/wiki/File:Neuristor.svg}

  Before delving into the architecture of our proposed AI accelerator, it is important to first examine the general structure of existing analog accelerators, which predominantly rely on non-volatile memory. At the heart of these accelerators lies the memristive crossbar, composed of non-volatile memory elements \cite{Xiao2020}. This crossbar structure facilitates highly parallelized in-memory computation, effectively minimizing data movement overhead. Figure \ref{fig:Crossbar} depicts a typical memristive crossbar, highlighting its fundamental design. These crossbars are widely utilized for both inference and training in neural networks, taking advantage of the analog domain to execute computations directly within the memory array.
  
  During inference, the crossbar computes a vector-matrix multiplication (VMM) between the input features ($\mathbf{x}^{(k)}$), applied as voltage signals at the rows of the crossbar, and the weights ($\mathbf{W}^{(k)}$), which are encoded as conductance values of the memristors. This operation is represented as:
  
  \begin{equation}\label{eq:VMM} \boldsymbol{\alpha}^{(k)} = \mathbf{W}^{(k)} \mathbf{x}^{(k)}. \end{equation}
  
  The output of the VMM, given in Equation \eqref{eq:VMM}, corresponds to the pre-activation output of a neuron in the current layer ($k$). The final output of the neuron is obtained by applying an activation function ($f$) to the result:
  
  \begin{equation}\label{eq:ForwardPass} \mathbf{x}^{(k+1)} = f\left(\boldsymbol{\alpha}^{(k)}\right). \end{equation}
  
  During training, the crossbar performs a matrix-vector multiplication (MVM) operation between the error signal ($\boldsymbol{\delta}^{(k)}$), now driven as currents on the columns of the crossbar, and the weights ($\mathbf{W}^{(k)}$). This operation, which is fundamental to the back-propagation process, is expressed as:
  
  \begin{equation}\label{eq:MVM} \boldsymbol{\epsilon}^{(k)} = \left(\mathbf{W}^{(k)}\right)^\mathrm{T} \boldsymbol{\delta}^{(k)}. \end{equation}
  
  The error is subsequently propagated to earlier layers using the following equation:
  
  \begin{equation}\label{eq:BackwardPass} \boldsymbol{\delta}^{(k-1)} = \boldsymbol{\epsilon}^{(k)} \odot f'\left(\boldsymbol{\alpha}^{(k-1)}\right), \end{equation}
  
  where $f'$ denotes the derivative of the activation function, and $\odot$ represents the element-wise product.
  
  This approach requires a significant digital overhead to compute activation functions and their gradients. Additionally, digital-to-analog converters (DACs) and analog-to-digital converters (ADCs) are necessary for generating and measuring the neural network signals. In large neural networks, signal repeaters are often required to compensate for voltage drops caused by long interconnects.
  
  Importantly, a critical challenge lies in the non-idealities of memristive devices, such as drift and cycle-to-cycle variability, which can significantly affect the accuracy and reliability of computations. These non-ideal characteristics complicate the integration of memristive devices into analog accelerators, as discussed comprehensively in \cite{Xiao2020}.
  
  In the remainder of this paper, we propose an analog accelerator that builds upon the standard crossbar architecture but instead leverages subthreshold MOS circuits. This approach addresses the challenges posed by non-volatile memory while improving overall performance and energy efficiency.

% \clearpage
\section{Theory}

This section introduces the mathematical formulation for solving stochastic gradient descent with L2 regularization (SGDr) in continuous time, establishing the theoretical foundation and guiding principles of our proposed analog accelerator. The SGDr can be expressed as the following discrete difference equation:

\begin{equation}\label{eq:dif} w[n+1] = w[n] - \alpha \delta[n] x[n] - \alpha \lambda w[n], \end{equation}

where $n$ denotes the discrete time step, $w$ represents the weight, $x$ is the input feature, $\delta[n]$ is the error term, $\alpha$ is the learning rate, and $\lambda$ is the regularization coefficient. Let $m$ denote the number of samples in the dataset; thus, one epoch is completed after processing $m$ samples.

\begin{figure}[h]
\centering
\includegraphics[width=0.3 \textwidth]{ 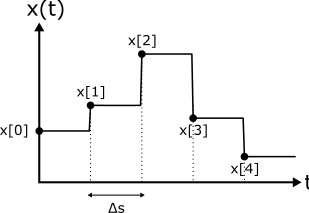}
\caption{ Represents the discrete-time input feature $x[n]$ as a continuous-time step function $x(t)$.} 
\label{fig:stepfunc}
\end{figure}

To approximate the difference equation \eqref{eq:dif} using a continuous differential equation, we represent $x[n]$ as a step function $x(t)$, defined as:

\begin{equation}\label{eq:step} x(t) = \sum_{n=0}^{\infty} x[n] \cdot \left[ u(t - (n+1)\Delta s) - u(t - n\Delta s) \right] \end{equation}

where $\Delta s$ denotes the hold time of each sample, and $u(t)$ is the unit step function. Figure \ref{fig:stepfunc} illustrates the relationship between the discrete signal $x[n]$ and its continuous counterpart $x(t)$. 
To model the behavior of a practical digital-to-analog converter (DAC), the rise time of each step is set to $\%0.5$ of $\Delta s$ for all results presented in this paper.

% \vspace{1cm}

Equation \eqref{eq:dif} can be approximated as the ordinary differential equation:

\begin{equation}\label{eq:diff}
\dot w(t)  + \frac{\alpha}{\Delta s} \lambda w(t) + \frac{\alpha}{\Delta s} \delta(t) x(t) = 0
\end{equation}

where $\dot w(t)$ represents the time derivative of $w(t)$. The detailed derivation of this equation is provided in Appendix \ref{proof_SGDCT}.

We refer to this differential equation \eqref{eq:diff} as the continuous-time stochastic gradient descent with L2-regularisation (SGDr-CT). The time required to complete a single epoch is  $ m \times \Delta s $, where $m$ is the number of samples in the dataset and $\Delta s$ is the hold time for each sample. Accordingly, the weights at the end of epoch $e$ can be sampled using the unit impulse function $\gamma(t)$ as follows:

\begin{equation}\label{eq:difff}
w[e] = w(t) \times \gamma (t- e \cdot m \cdot \Delta s)
\end{equation}

For efficient log-domain analog circuit realization, it is advantageous to use strictly positive (i.e., unidirectional) signals. This can be achieved by employing differential signaling to represent the model weights and inputs. Hence, we begin by modifying the SGDr-CT equation \eqref{eq:diff} for differential signaling before presenting the transistor-level implementation.

To simplify the derivation, we assume that the inputs are always strictly positive, whereas the weights (model parameters) are bidirectional. Let $w_i(t)$ be the difference between two strictly positive (i.e. uni-directional) signals, $w_i^+(t)$  and $w_i^-(t)$:

To simplify the derivation, we assume that the inputs are always strictly positive, whereas the weights (model parameters) can be bidirectional. Let $w_i(t)$ represent the difference between two strictly positive (i.e., unidirectional) signals, $w_i^+(t)$  and $w_i^-(t)$:

\begin{equation}\label{eq:w} w_{i}(t) = w_{i}^+(t) - w_{i}^-(t). \end{equation}

Similarly, we introduce $\delta^+(t)$ and $\delta^-(t)$  to represent  $\delta(t)$ as a differential signal. Using this formulation, the SGDr-CT equation \eqref{eq:diff} can be reformulated as an equality between two differential equations:

\begin{equation}\label{eq:tautologys}
\dot w_{i}^+(t)  + \frac{\alpha}{\Delta s} \lambda w_{i}^+(t) - \frac{\alpha}{\Delta s}  \delta^- x_i(t) = \dot w_{i}^-(t) + \frac{\alpha}{\Delta s} \lambda w_{i}^-(t) - \frac{\alpha}{\Delta s}  \delta^+ x_i(t) = c(t)
\end{equation}

where  \[ c(t) \in \mathbb{R} \]

Equation \eqref{eq:tautologys} is a tautology that holds true regardless of the value of $c(t)$.Therefore, we can separate \eqref{eq:tautologys} into two differential equations by setting $c(t)$ to $0$:

\begin{equation}\label{eq:pos}
\dot w_{i}^+(t)   + \frac{\alpha}{\Delta s} \lambda w_{i}^+(t) = \frac{\alpha}{\Delta s}  \delta^- x_i(t)
\end{equation}

\begin{equation}\label{eq:neg}
\dot w_{i}^-(t)   + \frac{\alpha}{\Delta s} \lambda w_{i}^-(t) = \frac{\alpha}{\Delta s}  \delta^+ x_i(t)
\end{equation}

We will refer to equation \eqref{eq:pos} as the positive learning equation and to equation \eqref{eq:neg} as the negative learning equation. This approach can be generalized to accommodate bidirectional inputs by expressing $x(t) = x^+(t) -  x^-(t)$, which results in four learning equations.

\section{Analog Circuits Realisation} \label{sa5}

\begin{figure}[!h]
    \centering
    \includegraphics[width = 0.90 \textwidth]{ 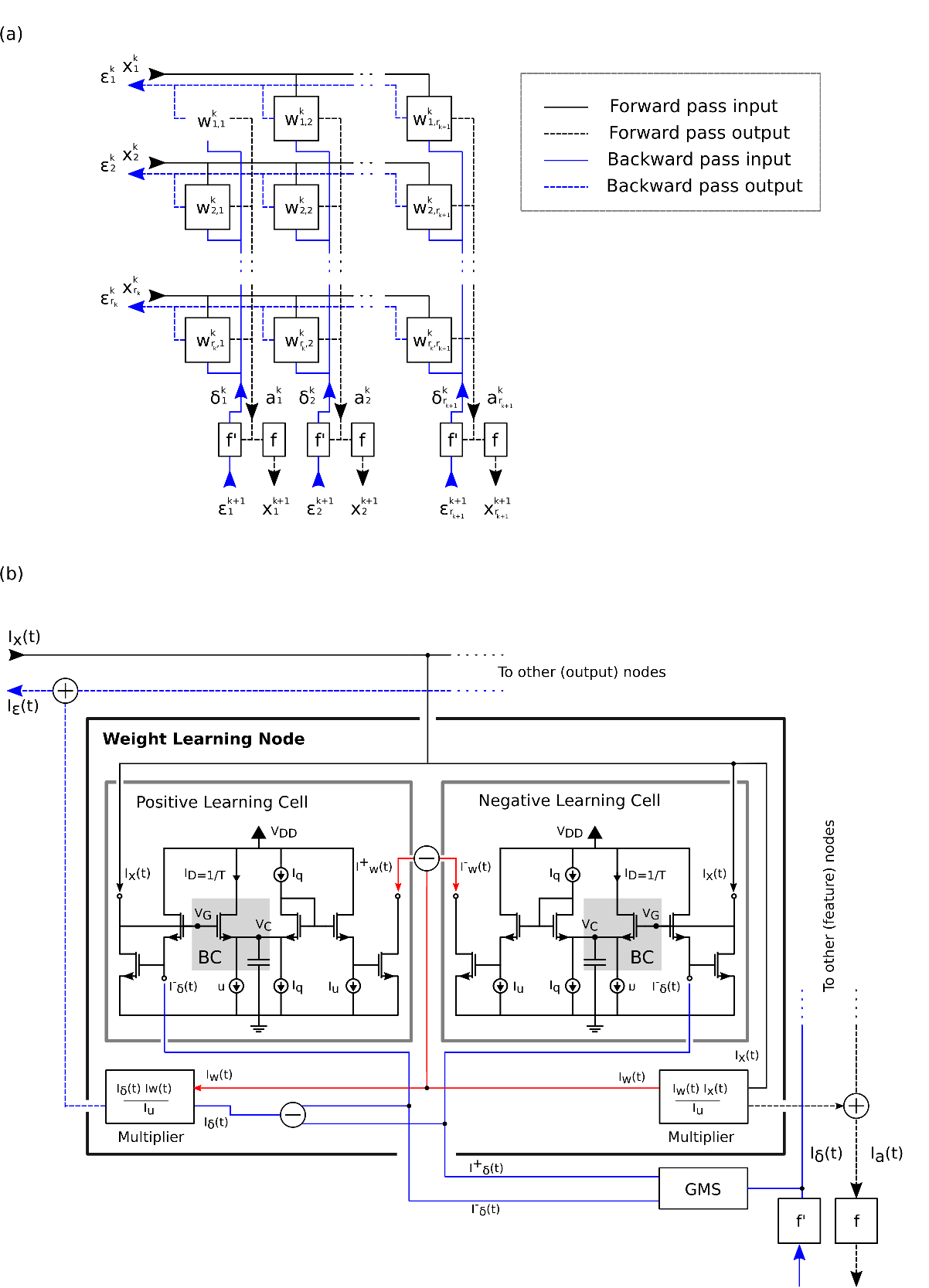}
    \caption{(a) Architecture of the proposed subthreshold CMOS analog accelerator, featuring a crossbar structure composed of weight-learning nodes. (b) Circuit-level realization of the weight-learning node, illustrating the integration of volatile memory and translinear MOS circuits for efficient weight updates.}
    \label{fig:Circuit}
\end{figure}

This section details the architecture and implementation of the proposed analog accelerator, which is based on well-characterized MOS log-domain circuits. The accelerator leverages the continuous-time stochastic gradient descent with L2-regularization (SGDr-CT) algorithm, described earlier, to enable efficient training of machine learning models in the continuous time domain.

The architecture of the proposed accelerator is illustrated in Figure \ref{fig:Circuit}. Similar to other analog accelerators in the literature, the design employs a crossbar array as its primary computational block. The crossbar is composed of weight-learning nodes, where each node is responsible for tuning or updating, in memory, a single weight in the neural network model being trained. The crossbar depicted in the figure represents a layer $k$, which contains $r_k$ inputs connected to $r_{k+1}$ outputs via the crossbar of the weight-learning nodes. These outputs subsequently serve as the inputs for the next layer ( $k+1$). Nodes in the same row are associated with the same feature (input), while nodes in the same column correspond to the same output. In smaller models, only a subset of these nodes is utilized; for example, a simple linear regression model may use only a single column.

In addition to the crossbar, the accelerator integrates analog blocks to realize the activation function ($f$) and its gradients ($f'$), both of which are essential for forward pass and back-propagation during training. The implementation of these blocks is particularly straightforward for the ReLU activation function, as detailed in section \ref{sa7}. Figure Figure \ref{fig:Circuit} highlights the datapath used for the forward pass and backward pass, respectively, highlighting the seamless integration of operations within the same structure.

As shown in Figure \ref{fig:Circuit}, the core of the architecture is the weight-learning node, which constitutes the fundamental building block of the crossbar. To better understand its functionality, we present its transistor-level implementation. The weight-learning node consists of two identical unidirectional cells that implement the positive \eqref{eq:pos} and negative \eqref{eq:neg} learning equations, respectively. These cells, referred to as the positive learning cell and negative learning cell, are illustrated in Figure \ref{fig:Circuit}.

Since the two cells are identical in design and operation, we focus our discussion on the positive learning cell. The mathematical representation of the positive learning cell is expressed as:

\begin{equation}\label{eq:plc}
\dot{I}_{w}^+(t) + \frac{u}{n C V_T} I_{w}^+(t) 
= \frac{I_q}{n C V_T} \frac{I_\delta^-(t) I_{x}(t)}{I_{u}},
\end{equation}

where \( \dot{I}_{w}^+(t) \) is the time derivative of the positive weight current, \( u \) is the unit constant, \( n \) is the subthreshold slope factor, \( C \) is the capacitance, \( V_T \) is the thermal voltage, \( I_q \) is the reference current, \( I_\delta^-(t) \) is the negative error current, \( I_{x}(t) \) is the input current, and \( I_{u} \) is the normalizing current. The full derivation of this equation is provided in Appendix \ref{drak} and relies on the Log-Domain State-Space approach and the Bernoulli Cell (BC) proposed in \cite{Drakakis1999}.

To realize the dynamics of the positive learning equation \eqref{eq:pos}, the mathematical variables are mapped into electrical currents within the circuit, as detailed in Table \ref{mapping1}. The bidirectional weight \( w(t) \), mapped to \( I_{w}(t) \), is computed as the difference between the currents \( I_{w}^+(t) \) and \( I_{w}^-(t) \), generated by the positive and negative learning cells, respectively.

Notably, the mapping relationship between the hyperparameters (e.g., \( \Delta s \) and \( \lambda \)) the circuit parameters exhibits a one-to-many cardinality. For instance, a desired learning rate ($\alpha$) can be obtained by adjusting the values of $I_q$ and/or $\Delta s$. Similarly, the regularisation coefficient ($\lambda$) can be tuned by adjusting $I_q$ and/or $u$. The primary advantage of this cardinality is that it enables using the same learning rate ($\alpha$) with smaller capacitance $C$, and hence smaller circuit footprint, simply by scaling down $\Delta s$.

% \vspace{1cm} %%%%%%%%%%%%%%%%%%%%%%%%%%
\begin{table}[!h]
%% increase table row spacing, adjust to taste
\renewcommand{\arraystretch}{1.5}
\caption{Mapping mathematical variables into physical current-mode signals. }
\label{mapping1}
\centering
\begin{tabular}{c|c}
\hline
Model Parameter  &  Electrical  Parameter \\
\hline
\hline
\\
 $w(t)$          &  $\displaystyle \frac{I^+_{w}(t)}{I_u}$ - $\displaystyle \frac{I^-_{w}(t)}{I_u}$ \\ \\
%  $w^+(t)$     /  $w^-(t)$       &  $\displaystyle \frac{I^+_{w}(t)}{I_u}$ / $\displaystyle \frac{I^-_{w}(t)}{I_u}$ \\ \\
% \hline \\
 $x(t)$            & $\displaystyle \frac{I_{x}(t)}{I_u}$ \\ \\
%\hline  \\
  $y(t)$            & $\displaystyle \frac{I_{y}(t)}{I_u}$ \\ \\
%\hline  \\
  $\delta(t)$       & $\displaystyle \frac{I^+_{\delta}(t)}{I_u}$ - $\displaystyle \frac{I^-_{\delta}(t)}{I_u}$\\ \\
  % $\delta^-(t)$ / $\delta^+(t)$        & $\displaystyle \frac{I^-_{\delta}(t)}{I_u}$ / $\displaystyle \frac{I^+_{\delta}(t)}{I_u}$\\ \\
%\hline \\  
 $\alpha$       & $\displaystyle \frac{I_q}{nCV_T}\Delta s$\\ \\
%   $\displaystyle\frac{\alpha}{\Delta s}$       & $\displaystyle \frac{I_q}{nCV_T}$\\ \\
%\hline \\
 $\lambda$      & $\displaystyle \frac{u}{I_q}$\\
 \\
 
\hline
\end{tabular}
\end{table}

In addition to the learning cells, the weight-learning node includes two multipliers, both implemented using translinear log-domain circuits.  The first multiplier computes the product of the weight and input currents during the forward pass, while the second multiplier second multiplier is employed during back-propagation to compute the product of the error and the weight currents. Finally, to support differential signaling of the error $\delta^(t)$ in back-propagation, the accelerator incorporates a geometric mean splitter (GMS) \cite{Yang2015}, which generates the differential current-mode error signals ($I^+_{\delta}(t)$ and $I^-_{\delta}(t)$).

In summary, similar to other analog accelerators in the literature, the proposed architecture employs a crossbar as its primary computational block. However, unlike traditional designs that utilize memristive elements, our approach introduces a novel weight-learning node based on volatile memory and translinear MOS circuits. This design leverages the energy efficiency of analog computation while utilizing well-characterized and mature circuit elements, ensuring both reliability and scalability.

% \clearpage
\section{Results}

In this section, the performance of the proposed analog accelerator is evaluated through two distinct experiments. First, we compare the accuracy of the circuit implementation (SGDr-CT) against the ideal SGDr algorithm on single-feature datasets. Second, we assess the ability of the circuit to fit a regression model, demonstrating its effectiveness in practical applications.

For all experiments presented in this section, the SGDr-CT circuit was implemented in AMS 0.35 µm CMOS technology and simulated using Spectre in Cadence Virtuoso. To provide a baseline for comparison, an ideal model implementing the SGDr algorithm was developed in Python.

\subsection{Single Neuron}

\begin{figure}[!ht]
\centering
\includegraphics[width = 0.8 \textwidth]{ 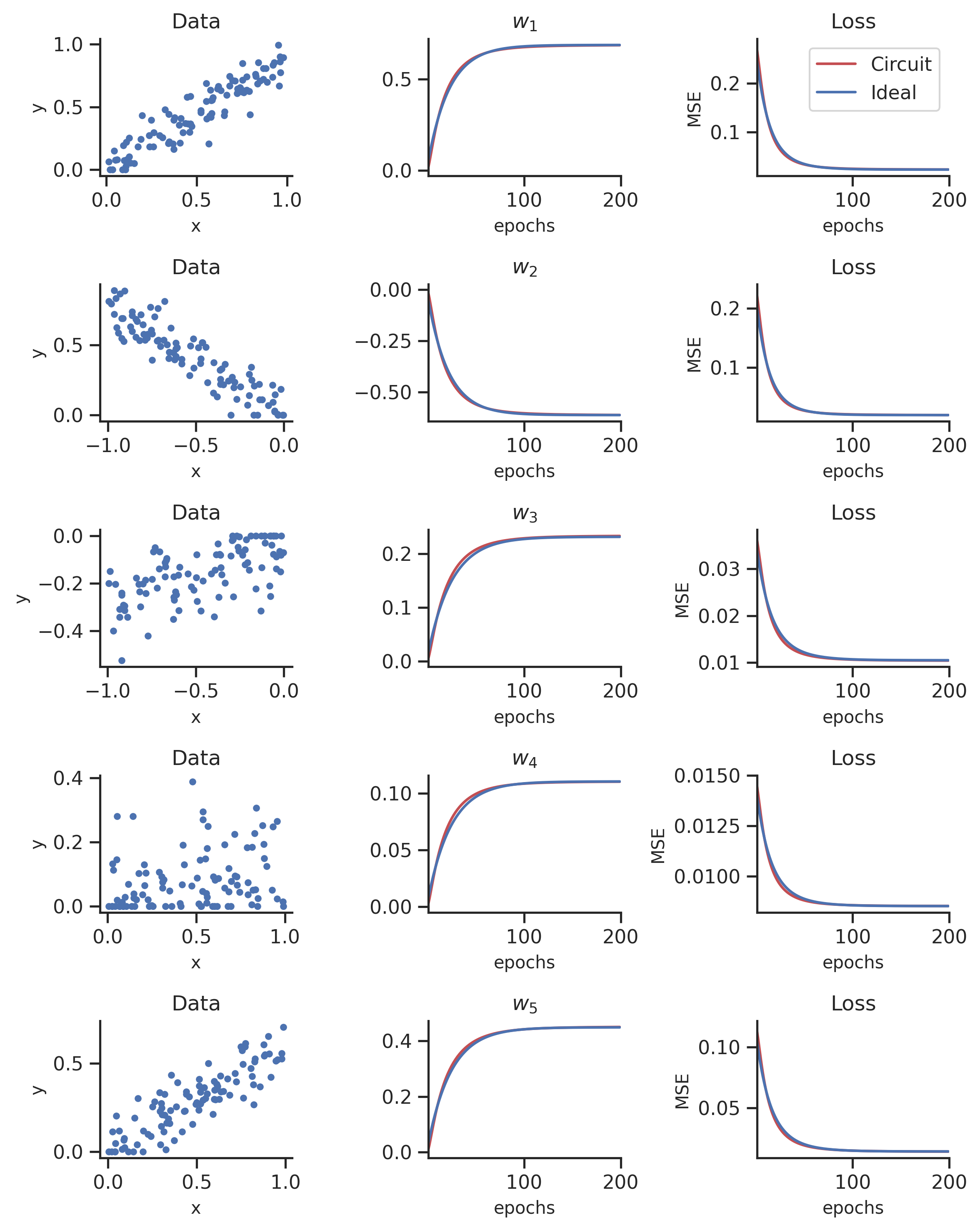}
\caption{The estimated weights and loss curves of the ideal SGDr model and the circuit implementation (SGDr-CT) for  five randomly generated datasets. The features were drawn from a uniform distribution (-1,1) with added Gaussian noise (scaled by 0.1 factor). The models were trained for 200 epochs with the following hyperparameters: $\lambda$ = 0.1, $\alpha$=$1e^{-3}$, $\Delta s$= 0.01 ms.}
\label{fig:DTvsCT}
\end{figure}

We evaluated our proposed analog AI accelerator architecture using five univariate datasets. Each dataset consisted of a randomly generated feature vector and weight pairings, sampled from a uniform distribution over \( (-1, 1) \). Gaussian noise, scaled by 0.1, was added to simulate realistic data. The distributions of the datasets are shown in Figure \ref{fig:DTvsCT}.

Both the circuit implementation (SGDr-CT) and the ideal mathematical implementation were used to fit the datasets. The regularization coefficient (\( \lambda \)) and learning rate (\( \alpha \)) were set to \( 1 \times 10^{-3} \) and 0.1, respectively, while the sampling time (\( \Delta s \)) was set to 0.01 milliseconds. Table \ref{nom2} provides the nominal values of the circuit parameters used to map the hyperparameters.

Figure \ref{fig:DTvsCT} compares the results, presenting the mean squared error (MSE) loss curve and the estimated weight over 200 epochs. The curves from the circuit and ideal model appear nearly identical, with maximum absolute errors of 0.71\% for the estimated weights and 0.87\% for the MSE values. Table \ref{tabSGD} provides a more detailed comparison of the results for both methods.

\begin{table}[!h]
%% increase table row spacing, adjust to taste
\renewcommand{\arraystretch}{2}
\caption{Nominal values for circuit parameters.}
\label{nom2}
\centering
\begin{tabular}{c|c}
\hline
Parameter &  Value\\
\hline
\hline

  $V_{DD}$       & 3.3 V  \\ 
  $C$       & 39 nF  \\ 
  $nV_T$       &  25.6 mV \\ 
% \hline \\
  $I_u$      &  10 nA\\ 
%\hline  \\
  $u$     &   10 nA \\ 
%\hline  \\
  $I_q$        &  100 nA\\ 
%\hline \\  
  $\displaystyle\frac{W}{L}$ & $\displaystyle \frac{1  \mu m}{0.5 \mu m}$\\ \\

\hline
\end{tabular}
\end{table}

\begin{table}[!h]
\renewcommand{\arraystretch}{2}
\caption{Comparing the ideal and circuit implementations on five random datasets. All values reported are sampled after the 200th epoch.}
\label{tabSGD}
\centering

\begin{tabularx}{0.8 \textwidth}{ 
>{\centering\arraybackslash}XX
>{\centering\arraybackslash}X
>{\centering\arraybackslash}X
>{\centering\arraybackslash}X}

\hline
\multicolumn{2}{>{\centering\arraybackslash}X}{Dataset\#}             & Ideal & Circuit &  Absolute Error (\%) \\   
\hline
\hline

\multicolumn{1}{>{\centering\arraybackslash}X}{\multirow{2}{*}{1}} &  Weight & 0.687 & 0.686 & 0.184\% \\
\multicolumn{1}{>{\centering\arraybackslash}X}{}                   &  MSE      & 0.023 & 0.023 & 0.749\% \\
\hline

\multicolumn{1}{>{\centering\arraybackslash}X}{\multirow{2}{*}{1}} &  Weight & -0.611 & -0.611 & 0.108\% \\
\multicolumn{1}{>{\centering\arraybackslash}X}{}                   &  MSE      & 0.020 & 0.020 & 0.403\% \\
\hline

\multicolumn{1}{>{\centering\arraybackslash}X}{\multirow{2}{*}{1}} &  Weight & 0.231 & 0.233 & 0.706\% \\
\multicolumn{1}{>{\centering\arraybackslash}X}{}                   &  MSE      & 0.010 & 0.010 & 0.712\% \\
\hline

\multicolumn{1}{>{\centering\arraybackslash}X}{\multirow{2}{*}{1}} &  Weight & 0.111 & 0.110 & 0.174\% \\
\multicolumn{1}{>{\centering\arraybackslash}X}{}                   &  MSE      & 0.009 & 0.009 & 0.050\% \\
\hline

\multicolumn{1}{>{\centering\arraybackslash}X}{\multirow{2}{*}{1}} &  Weight & 0.449 & 0.450 & 0.310\% \\
\multicolumn{1}{>{\centering\arraybackslash}X}{}                   &  MSE      & 0.014 & 0.014 & 0.868\% \\
\hline

\end{tabularx}

\end{table}

\begin{figure}[!h]
\centering
\includegraphics[width = 0.8 \textwidth]{ 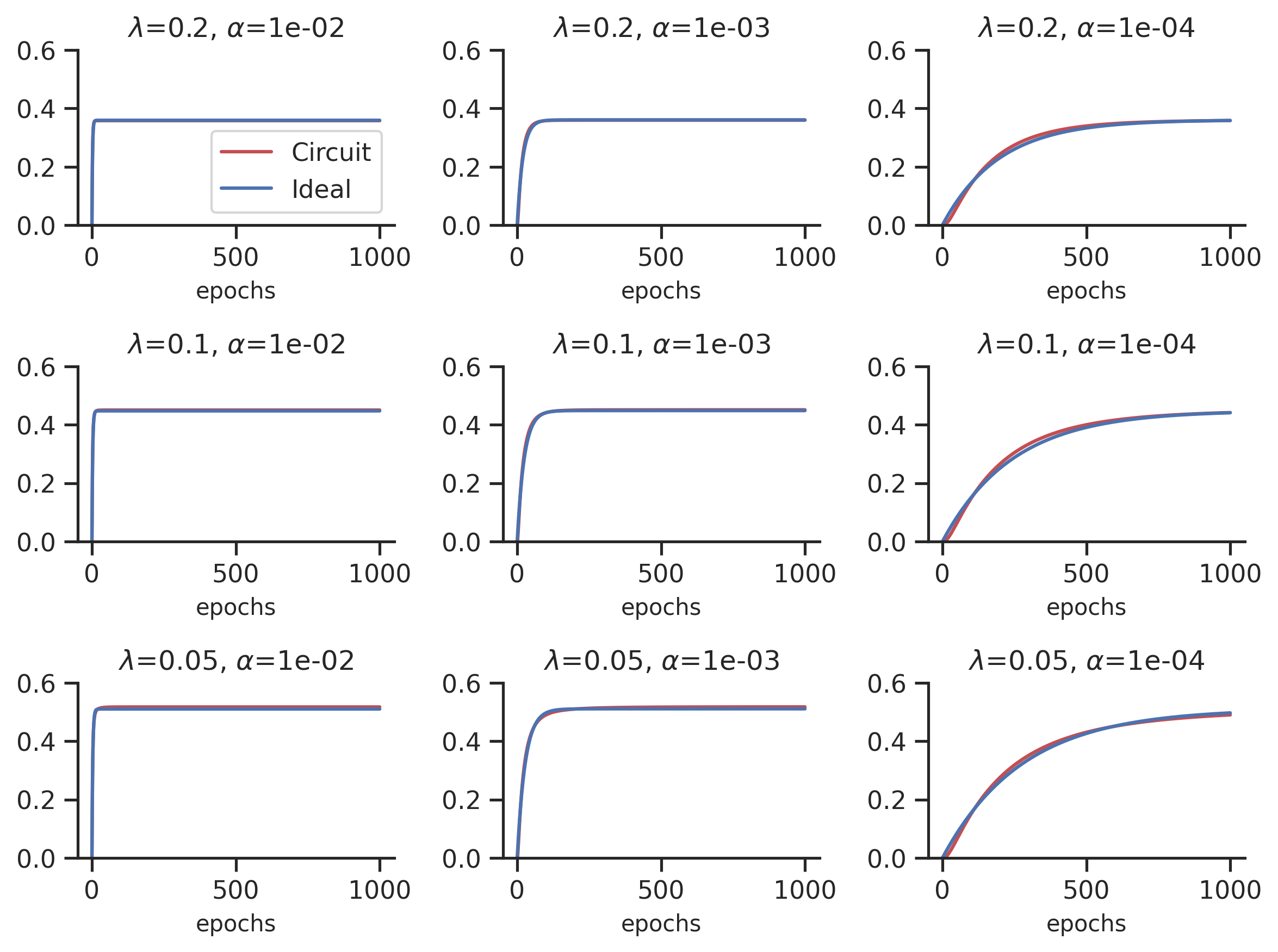}
\caption{Comparison of the accuracy of the ideal and circuit models for various hyperparameter values. The learning rate (\( \alpha \)) was set to \( 1 \times 10^{-2} \), \( 1 \times 10^{-3} \), and \( 1 \times 10^{-4} \), while the regularization coefficient (\( \lambda \)) was set to 0.2, 0.1, and 0.05.}
\label{fig:circVsIdeal2}
\end{figure}

Next, we examine the accuracy of the circuit in supporting a wide range of hyperparameter values (\( \lambda \) and \( \alpha \)). Starting with the previous nominal values, the learning rate was set to \( \alpha \) to $1e^{-2}$, $1e^{-3}$, and $1e^{-4}$ by varying \( \Delta s \). Simulations were repeated for \( \lambda \) values of 0.2, 0.1, and 0.05. Figure \ref{fig:circVsIdeal2} illustrates the results.

For \( \lambda = 0.2 \) and \( \lambda = 0.05 \), slight deviations were observed between the circuit and the ideal behavior. These deviations were corrected by tuning the mapped currents in the circuit to 0.27 and 0.03, respectively. The deviations occurred because the transistor dimensions were optimized for \( \lambda = 0.1 \). This tuning process aligns with the standard practice of hyperparameter tuning in the machine learning workflow, and further discussion is provided in Section \ref{sa7}.

In summary, these results demonstrate that the performance of the circuit implementation of SGDr-CT closely matches that of the ideal SGDr model. The negligible differences confirm the accuracy and reliability of the proposed analog accelerator for stochastic gradient descent-based training.
    
\clearpage
\subsection{Model Fitting}

The proposed analog AI accelerator was evaluated by training a linear regression model on the Boston Housing dataset \cite{Harrison1978}, a widely utilized benchmark for assessing machine learning algorithms \cite{8556738,fairlearn_boston}. The dataset comprises 506 samples, with 13 features representing various socioeconomic and geographical factors, used to predict the median home value in neighborhoods of Boston, Massachusetts. The Boston Housing dataset has been critiqued for embedding biases, particularly through the inclusion of a variable engineered under the assumption that racial self-segregation positively impacts house prices \cite{scikit-learn_boston, fairlearn_boston}. While this dataset was chosen for historical benchmarking purposes \cite{8556738,fairlearn_boston}, future work will explore alternative datasets that promote fairness in AI model evaluation.

Prior to training, the dataset was preprocessed to ensure uniform scaling and improve model training. Features were normalized using max normalization, and their means were shifted to 0.8 through the application of appropriate offsets. The dataset was subsequently divided into training and testing subsets, with 80\% of the data allocated for training and the remaining 20\% reserved for testing.

\begin{figure}[!ht]
\centering
\includegraphics[width = 0.8 \textwidth]{ 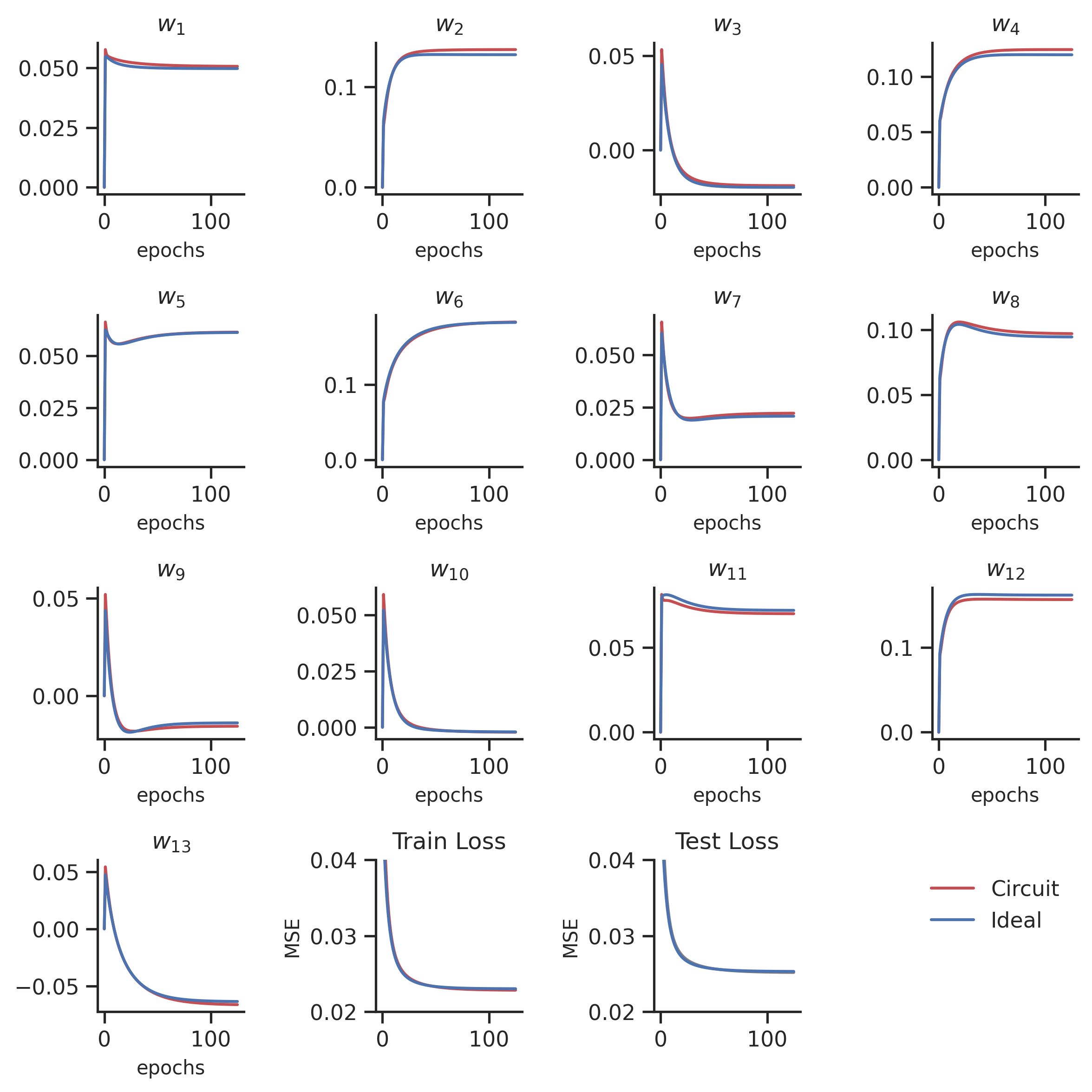}
\caption{Estimated weights and loss curves of the ideal model and circuit for fitting linear regression on the Boston Housing dataset. The models were trained for 125 epochs using the following hyperparameter values: $\lambda$ = 0.1, $\alpha$=$1e^{-3}$, $\Delta s$= 0.01 ms.}
\label{fig:Boston3}
\end{figure}

Figure \ref{fig:Boston3} illustrates the predicted weights obtained from the circuit implementation and the ideal mathematical model. The figure also presents the training and test loss curves. Overall, the predictions from the circuit closely align with those of the ideal model. A quantitative summary of the results at the final epoch is provided in Table \ref{bbbbbb}. As shown, the maximum error, expressed as the absolute difference between the circuit and ideal predictions, is 0.527\%. This error corresponds to an 8-bit resolution for weight values with a full-scale range of \([-1, 1]\). The accuracy in terms of the number of bits is determined using the following equation:

\begin{equation}\label{eq:bitsnum}
\text{\#Bits} = \floor*{ - \log_2{ \left( \frac{\text{Max error}}{\text{Full scale}} \right) } }
\end{equation}

Notably, the training and test loss curves align more closely with the ideal behavior compared to the parameter curves, with percentage errors remaining below 1\%. This observation suggests that the accelerator effectively compensates for circuit non-idealities, converging to a locally optimal solution with a similar loss to the ideal model, albeit with slight differences in the parameter values.

% \vspace{1cm}
\begin{table}[!h]
\renewcommand{\arraystretch}{2}
\caption{Comparison of the accuracy between the ideal and circuit models on the Boston Housing dataset. The error represents the percentage absolute difference between the circuit and the ideal model. All values are reported from the final epoch.}

\label{bbbbbb}
\centering
\begin{tabularx}{0.8 \textwidth}{ 
>{\centering\arraybackslash}X
>{\centering\arraybackslash}X
>{\centering\arraybackslash}X
>{\centering\arraybackslash}X}

\hline
Parameter &  Ideal  &  Circuit &   Absolute Error (\%) \\
\hline
\hline
$w_{1}$  & 0.050  & 0.051  & 0.093\% \\
$w_{2}$  & 0.132  & 0.137  & 0.508\% \\
$w_{3}$  & -0.020 & -0.019 & 0.089\% \\
$w_{4}$  & 0.120  & 0.124  & 0.470\% \\
$w_{5}$  & 0.061  & 0.061  & 0.017\% \\
$w_{6}$  & 0.182  & 0.183  & 0.054\% \\
$w_{7}$  & 0.021  & 0.022  & 0.142\% \\
$w_{8}$  & 0.095  & 0.097  & 0.252\% \\
$w_{9}$  & -0.014 & -0.016 & 0.172\% \\
$w_{10}$ & -0.002 & -0.002 & 0.015\% \\
$w_{11}$ & 0.072  & 0.070  & 0.198\% \\
$w_{12}$ & 0.162  & 0.157  & \underline{\textbf{0.527\%}} \\
$w_{13}$ & -0.064 & -0.066 & 0.276\% \\
\hline
Training Loss  & 0.023  & 0.023  & 0.019\% \\
Testing Loss   & 0.025  & 0.025  & 0.014\% \\
 
% % \hline 

%  $\lambda$      & 0.1 \\ \\
%   $\alpha$      & $1e^{-3}$ \\ \\
%   $\Delta s$      & 0.01 ms \\ \\

\hline
\end{tabularx}
\end{table}

% \clearpage
\section{Discussion and Future Work} \label{sa7}

This paper has focused on validating the core building block of the proposed analog accelerator: the weight-learning node. While the results demonstrate the feasibility and accuracy of the weight update mechanism, several design considerations remain critical for the complete realization of the accelerator. These considerations are discussed below.

\subsection{Inclusion of Bias Term}
The proposed design effectively implements the weight update rule; however, the realization of the bias term has not been addressed in this work. The inclusion of a bias term is essential for training machine learning models, as it enhances their ability to fit data accurately by shifting the decision boundary.

The continuous-time differential equation governing the update of the bias term is expressed as:

\begin{equation}\label{eq:bias}
\dot{w}_0(t) = - \frac{\alpha}{\Delta s} \delta(t),
\end{equation}

where \( w_0(t) \) denotes the bias term, \( \alpha \) is the learning rate, \( \Delta s \) is the sampling time, and \( \delta(t) \) represents the error signal.

The update rule for the bias term, as described in \eqref{eq:bias}, is straightforward and can be efficiently implemented using a Seevinck integrator. The Seevinck integrator \cite{Gerosa2004,Seevinck1990} is a fully differential log-domain integrator that aligns well with the analog nature of the proposed accelerator. Its capability to handle differential signals and its compatibility with log-domain circuits make it a suitable choice for realizing the bias term update.

The integration of the bias term into the accelerator design will enable a more comprehensive evaluation of the accelerator performance across diverse machine learning tasks, particularly those requiring complex decision boundaries. This enhancement represents a natural extension of the current work and will be a focus of future investigations.

\subsection{Regularization Coefficient}

The proposed accelerator was shown to support regularization coefficients (\( \lambda \)) as small as 0.05. However, since \( \lambda \) is represented as the ratio of two currents (\( u/I_q \)), smaller values of \( \lambda \) values would result in transistors operating outside the weak inversion region.

\begin{figure}[!h]
  \centering
  \includegraphics[width= 0.6 \textwidth]{ 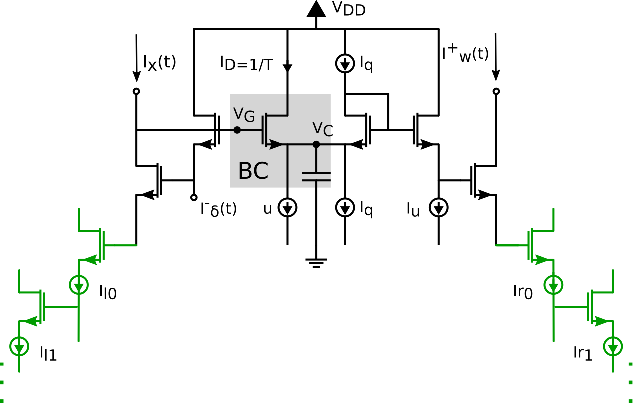}
  \caption{Extending the translinear loop in order to support smaller values for the regularisation coefficient $\lambda$.}
  \label{fig:extendingloop}
\end{figure}

To address this limitation, the translinear loop can be extended by stacking additional transistors, as illustrated in green in Figure \ref{fig:extendingloop}. This modification ensures that the circuit operates entirely within the subthreshold region, even for very small \( \lambda \) values. The updated mathematical equation governing the extended circuit is expressed as:

\begin{equation}\label{eq:circuitexp}
\dot{I}_{w}^+(t) + \frac{u}{nCV_T} I_{w}^+(t) = \frac{\prod_{i} I_{li}}{\prod_{j} I_{ri}} \frac{I_q}{nCV_T} \frac{I_\delta^-(t) I_{z}(t)}{I_{u}},
\end{equation}

where \( I_{li} \) and \( I_{ri} \) denote the left and right currents in the extended translinear loop, respectively.

The hyperparameters can be mapped to electrical currents as detailed in Table \ref{mappinglamda}. To support regularization coefficients two orders of magnitude smaller (i.e., \( 0.05 \times e^{-2} \)), two additional transistors can be added to each side of the translinear loop, with the bias currents set as \( I_{r0} = I_{r1} = 10 \times I_{l0} = 10 \times I_{l1} \).

% ($5e^{-4}$)
\begin{table}[!h]
\renewcommand{\arraystretch}{1.5}
\caption{Modified mapping of hyperparameters to physical circuit parameters.}
\label{mappinglamda}
\centering
\begin{tabular}{c|c}
\hline
Model Parameter  &  Electrical  Parameter \\
\hline
\hline
\\
$\displaystyle\alpha$    &    $\displaystyle \frac{\prod_{i} I_{li}}{\prod_{j}  I_{ri}} \frac{I_q}{nCV_T} \Delta s$
 \\  \\
 $\lambda$ &     $\displaystyle \frac{\prod_{j}  I_{ri}}{\prod_{i} I_{li}} \frac{u}{I_q}$  \\  \\
 
\hline
\end{tabular}
\end{table}

\subsection{ReLU Activation Function}
The ReLU activation function is widely employed in AI/ML applications due to its simplicity and effectiveness. Conveniently, both the ReLU function and its derivative can be efficiently implemented in silicon using analog circuits. The ReLU function can be realized with a current mirror, where the diode-connected transistor inherently permits only positive values (unidirectional currents) to pass. Similarly, the derivative of the ReLU function can be effectively implemented using a current comparator based on the Winner-Take-All (WTA) topology \cite{Moro-Fria2011}.

\subsection{Process Nodes}
As discussed earlier, a key advantage of the one-to-many mapping between hyperparameters and circuit parameters is the ability to scale the chip area by using smaller capacitance (\( C \)) through the reduction of \( \Delta s \). This not only decreases the circuit area but also increases throughput and reduces training time. Therefore, the proposed architecture scales effectively with faster process nodes. Future work should focus on realizing the accelerator using advanced fabrication technologies.

\appendix
% \clearpage
\section{Mathematical Proof} \label{ch:AppA}

\subsection{SGDr-CT Formulation} \label{proof_SGDCT}

\begin{proof}

Consider the first-order ordinary differential equation:

\begin{equation}\label{eq:S001}
\dot{w}(t) + \frac{\alpha}{\Delta s} \lambda w(t) + \frac{\alpha}{\Delta s} \delta(t) x(t) = 0
\end{equation}

To derive the discrete-time formulation, integrate \eqref{eq:S001} with respect to \( t \) over the interval \([n \Delta s, n \Delta s + \Delta s]\):

\begin{equation}\label{eq:S002}
w((n+1) \Delta s) - w(n \Delta s) = - \frac{\alpha}{\Delta s} \int_{n \Delta s}^{(n+1) \Delta s} [\lambda w(t) + \delta(t) x(t)] \, dt
\end{equation}

Assuming that \( \delta(t) \) and \( w(t) \) vary minimally over the interval \( \Delta s \), and applying Euler’s method, we approximate:

\begin{equation}\label{eq:S003}
  w((n+1) \Delta s)  \approx  w(n \Delta s) - \frac{\alpha}{\Delta s} \Delta s  [\lambda w(n \Delta s) +  \delta(n \Delta s) x(n \Delta s) ] 
\end{equation}
  
\begin{equation}\label{eq:S004}
  \leadsto \,\, w((n+1) \Delta s) = w(n \Delta s) - \alpha [\lambda w(n \Delta s) +  \delta(n \Delta s) x(n \Delta s) ] 
\end{equation}

Since \( x(n \Delta s) \) corresponds to the discrete-time variable \( x[n] \) (as illustrated in Figure \ref{fig:stepfunc}), \eqref{eq:S004} can be rewritten in its discrete-time form as the stochastic gradient descent (SGD) difference equation:

\begin{equation}\label{eq:S005}
w[n+1] = w[n] - \alpha \delta[n] x[n] - \alpha \lambda w[n]
\end{equation}

This establishes the equivalence between the continuous-time SGDr-CT formulation and the discrete-time SGDr difference equation.
\end{proof}

\subsection{Positive Learning Cell Expression} \label{drak}

\begin{proof}

We begin by analyzing the Bernoulli Cell (BC) in the positive learning cell, followed by an examination of its translinear loop.

% \subsubsection{The Bernoulli Cell (BC)}

The drain current of the NMOS transistor within the Bernoulli Cell (BC) of the positive learning cell can be expressed as follows:

\begin{equation}\label{eq:S01}
{I}_{D}(t) = u + i_C(t)
\end{equation}

where $i_C(t)$ represents the current flowing through a capacitor with capacitance $C$:

\begin{equation}\label{eq:S02}
i_C(t) = C \dot{V}_C(t)
\end{equation}

In the subthreshold region of operation, the drain current of the NMOS transistor can be approximated as:

\begin{equation}\label{eq:S03}
I_D(t) = I_{D0} \cdot e^{\displaystyle\frac{ V_G(t) - V_C(t)}{n V_t}} 
\end{equation}

where \( I_{D0} \) is a constant that depends on the technology and the transistor aspect ratio. Differentiating equation \eqref{eq:S01} yields the following expression:

\begin{equation}\label{eq:S05}
\dot{I}_D(t) = (\dot{V}_G(t) - \dot{V}_C(t)) \frac{{I}_D(t)}{n V_t} \,\,\, \leadsto \,\,\,  \dot{V}_C(t)  = \dot{V}_G(t) - n V_t \frac{\dot{I}_D(t)}{{I}_D(t)} 
\end{equation}

From equations \eqref{eq:S01}, \eqref{eq:S02}, and \eqref{eq:S05}, the following non-linear differential equation is obtained:

\begin{equation}\label{eq:S07}
{I}_{D}(t) =  C\dot{V}_G(t) + u -  nCV_t \frac{\dot{I}_D(t)}{{I}_D(t)} 
\end{equation}

\begin{equation}\label{eq:S09}
\leadsto \,\,\,  \dot{I}_D(t) = ( \frac{\dot{V}_G(t)}{nV_t} + \frac{u}{nCV_t}) {I}_D(t) -  \frac{{I}^2_D(t)} {nCV_t} 
\end{equation}

The above equation is in Bernoulli form. To linearize it, we substitute a new variable \( T(t) \), defined as \( T(t) = \frac{1}{I_D} \). Taking the time derivative of \( T(t) \), we obtain $\dot{T}(t) = -\dot{I}_D(t) \cdot T^2(t)$. Substituting this relationship into the original equation yields:

\begin{equation}\label{eq:S10}
 - \frac{\dot{T(t)}}{T^2(t)}  = ( \frac{\dot{V}_G(t)}{nV_t} + \frac{u}{nCV_t}) \frac{1}{T(t)} -  \frac{1} {nCV_t \cdot T^2(t)} 
\end{equation}

\begin{equation}\label{eq:S11}
 \leadsto \,\,\, \dot{T(t)} + ( \frac{\dot{V}_G(t)}{nV_t} + \frac{u}{nCV_t}) T(t) =  \frac{1} {nCV_t} 
\end{equation}

This transformation results in a linear differential equation. Next, we derive an expression for \( \dot{V}_G(t) \) as a function of the input currents, specifically \( I_\delta^-(t) \) and \( I_z(t) \).

The gate voltage of the transistor (denoted as \( V_G \) in the circuit diagram) can be expressed in terms of the input currents as follows:

\begin{equation}\label{eq:S12}
{V}_G(t) =   nV_t \ln(\frac{I_\delta^-(t)}{I_S}) + nV_t \ln(\frac{I_z(t)}{I_S})
\end{equation}

Differentiating the above equation results in:

\begin{equation}\label{eq:S13}
\dot{V}_G(t) =  nV_t \frac{\dot{I}_\delta^-(t)}{I_\delta^-(t)} +  nV_t \frac{\dot{I}_z(t)}{{I}_z(t)} \end{equation}

From equations \eqref{eq:S11} and \eqref{eq:S13}, and dividing by \( T(t) \), we obtain:

\begin{equation}\label{eq:S14}
 \frac{\dot{T(t)}}{T(t)} + \frac{\dot{I}_\delta^-(t)}{I_\delta^-(t)} +  \frac{\dot{I}_z(t)}{{I}_z(t)}  + \frac{u}{nCV_t}  =  \frac{1} {nCV_t T(t)} \end{equation}

\begin{equation}\label{eq:S15}
\leadsto \,\,\,  \frac{d}{dt} \ln[  T(t) \cdot I_\delta^-(t) \cdot {I}_z(t)      ]+   \frac{u}{nCV_t}  =  \frac{1} {nCV_t T(t)} \end{equation}

Substituting a new variable \( \omega(t) \), defined as \( \omega(t) = T(t) \cdot I_\delta^-(t) \cdot I_z(t) \) (note that \( \omega(t) \) should not be confused with \( w(t) \), which represents the weights in the neural network), we obtain:

\begin{equation}\label{eq:S16}
\frac{\dot{\omega}(t)}{\omega(t)} +   \frac{u}{nCV_t}  =  \frac{ I_\delta^-(t) \cdot {I}_z(t) } {nCV_t \cdot \omega(t)} \end{equation}

\begin{equation}\label{eq:S17}
\leadsto \,\,\, nCV_t \dot{\omega}(t) +   u \cdot \omega(t)  =  I_\delta^-(t)   \cdot {I}_z(t)  \end{equation}
 
The variable \( \omega \) can be expressed as a function of the output current using the translinear principle. By analyzing the translinear loop in the positive learning cell circuit, the following relationship is derived:
 
\begin{equation}\label{eq:S18}
I_z(t) \cdot I_\delta^-(t) \cdot I_q = \frac{1}{T(t)} \cdot I_u \cdot I_w^+(t),
\end{equation}

Since $\omega(t)$ = $ T(t) \cdot I_\delta^-(t) \cdot {I}_z(t) $, we obtain:

\begin{equation}\label{eq:S19}
\omega(t) = \frac{I_u \cdot I_w^+(t)}{I_q} \,\,\, \leadsto \,\,\, \dot{\omega}(t) = \frac{I_u \cdot \dot{I}_w^+(t)}{I_q}.
\end{equation}

Finally, combining \eqref{eq:S18} and \eqref{eq:S19}, the following expression is derived:

\begin{equation}\label{eq:S20}
nC V_T \cdot \frac{I_u \cdot \dot{I}_w^+(t)}{I_q} + u \cdot \frac{I_u \cdot I_w^+(t)}{I_q} = I_\delta^-(t) \cdot I_z(t).
\end{equation}

Reorganizing terms, we arrive at:

\begin{equation}\label{eq:S21}
\dot{I}_w^+(t) + \frac{u}{nC V_T} I_w^+(t) = \frac{I_q}{nC V_T} \frac{I_\delta^-(t) I_z(t)}{I_u},
\end{equation}

which is the governing expression of the positive learning cell.
 
\end{proof}

%Bibliography
\bibliographystyle{unsrt}  
\bibliography{references}  

\begin{thebibliography}{10}

\bibitem{Gao2019}
Dashan Gao, Yang Liu, Anbu Huang, Ce~Ju, Han Yu, and Qiang Yang.
\newblock Privacy-preserving heterogeneous federated transfer learning.
\newblock In {\em 2019 IEEE International Conference on Big Data (Big Data)},
  pages 2552--2559, 2019.

\bibitem{NIPS2017_3f5ee243}
Ashish Vaswani, Noam Shazeer, Niki Parmar, Jakob Uszkoreit, Llion Jones,
  Aidan~N Gomez, \L~ukasz Kaiser, and Illia Polosukhin.
\newblock Attention is all you need.
\newblock In I.~Guyon, U.~Von Luxburg, S.~Bengio, H.~Wallach, R.~Fergus,
  S.~Vishwanathan, and R.~Garnett, editors, {\em Advances in Neural Information
  Processing Systems}, volume~30. Curran Associates, Inc., 2017.

\bibitem{Coates2013}
Adam Coates, Brody Huval, Tao Wang, David~J. Wu, Andrew~Y. Ng, and Bryan
  Catanzaro.
\newblock Deep learning with cots hpc systems.
\newblock In {\em Proceedings of the 30th International Conference on
  International Conference on Machine Learning - Volume 28}, ICML'13, page
  III–1337–III–1345. JMLR.org, 2013.

\bibitem{anthony2020carbontracker}
Lasse F.~Wolff Anthony, Benjamin Kanding, and Raghavendra Selvan.
\newblock Carbontracker: Tracking and predicting the carbon footprint of
  training deep learning models.
\newblock ICML Workshop on Challenges in Deploying and monitoring Machine
  Learning Systems, July 2020.
\newblock arXiv:2007.03051.

\bibitem{Andri2016}
Renzo Andri, Lukas Cavigelli, Davide Rossi, and Luca Benini.
\newblock Yodann: An ultra-low power convolutional neural network accelerator
  based on binary weights.
\newblock In {\em 2016 IEEE Computer Society Annual Symposium on VLSI
  (ISVLSI)}, pages 236--241, 2016.

\bibitem{strubell-etal-2019-energy}
Emma Strubell, Ananya Ganesh, and Andrew McCallum.
\newblock Energy and policy considerations for deep learning in {NLP}.
\newblock In Anna Korhonen, David Traum, and Llu{\'\i}s M{\`a}rquez, editors,
  {\em Proceedings of the 57th Annual Meeting of the Association for
  Computational Linguistics}, pages 3645--3650, Florence, Italy, July 2019.
  Association for Computational Linguistics.

\bibitem{Xiao2020}
T.~Patrick Xiao, Christopher~H. Bennett, Ben Feinberg, Sapan Agarwal, and
  Matthew~J. Marinella.
\newblock {Analog architectures for neural network acceleration based on
  non-volatile memory}.
\newblock {\em Applied Physics Reviews}, 7(3), 2020.

\bibitem{Sze2017}
Vivienne Sze, Yu~Hsin Chen, Tien~Ju Yang, and Joel~S. Emer.
\newblock {Efficient Processing of Deep Neural Networks: A Tutorial and
  Survey}.
\newblock {\em Proceedings of the IEEE}, 105(12):2295--2329, 2017.

\bibitem{Reuther2020}
Albert Reuther, Peter Michaleas, Michael Jones, Vijay Gadepally, Siddharth
  Samsi, and Jeremy Kepner.
\newblock {Survey of Machine Learning Accelerators}.
\newblock {\em 2020 IEEE High Performance Extreme Computing Conference, HPEC
  2020}, pages 1--9, 2020.

\bibitem{Chung2018}
Eric Chung, Jeremy Fowers, Kalin Ovtcharov, Michael Papamichael, Adrian
  Caulfield, Todd Massengill, Ming Liu, Daniel Lo, Shlomi Alkalay, Michael
  Haselman, Maleen Abeydeera, Logan Adams, Hari Angepat, Christian Boehn, Derek
  Chiou, Oren Firestein, Alessandro Forin, Kang~Su Gatlin, Mahdi Ghandi,
  Stephen Heil, Kyle Holohan, Ahmad El~Husseini, Tamas Juhasz, Kara Kagi,
  Ratna~K. Kovvuri, Sitaram Lanka, Friedel van Megen, Dima Mukhortov, Prerak
  Patel, Brandon Perez, Amanda Rapsang, Steven Reinhardt, Bita Rouhani, Adam
  Sapek, Raja Seera, Sangeetha Shekar, Balaji Sridharan, Gabriel Weisz, Lisa
  Woods, Phillip Yi~Xiao, Dan Zhang, Ritchie Zhao, and Doug Burger.
\newblock Serving dnns in real time at datacenter scale with project brainwave.
\newblock {\em IEEE Micro}, 38(2):8--20, 2018.

\bibitem{Jouppi2017}
Norman~P. Jouppi, Cliff Young, Nishant Patil, David Patterson, Gaurav Agrawal,
  Raminder Bajwa, Sarah Bates, Suresh Bhatia, Nan Boden, Al~Borchers, Rick
  Boyle, Pierre luc Cantin, Clifford Chao, Chris Clark, Jeremy Coriell, Mike
  Daley, Matt Dau, Jeffrey Dean, Ben Gelb, Tara~Vazir Ghaemmaghami, Rajendra
  Gottipati, William Gulland, Robert Hagmann, C.~Richard Ho, Doug Hogberg, John
  Hu, Robert Hundt, Dan Hurt, Julian Ibarz, Aaron Jaffey, Alek Jaworski,
  Alexander Kaplan, Harshit Khaitan, Daniel Killebrew, Andy Koch, Naveen Kumar,
  Steve Lacy, James Laudon, James Law, Diemthu Le, Chris Leary, Zhuyuan Liu,
  Kyle Lucke, Alan Lundin, Gordon MacKean, Adriana Maggiore, Maire Mahony,
  Kieran Miller, Rahul Nagarajan, Ravi Narayanaswami, Ray Ni, Kathy Nix, Thomas
  Norrie, Mark Omernick, Narayana Penukonda, Andy Phelps, Jonathan Ross, Matt
  Ross, Amir Salek, Emad Samadiani, Chris Severn, Gregory Sizikov, Matthew
  Snelham, Jed Souter, Dan Steinberg, Andy Swing, Mercedes Tan, Gregory
  Thorson, Bo~Tian, Horia Toma, Erick Tuttle, Vijay Vasudevan, Richard Walter,
  Walter Wang, Eric Wilcox, and Doe~Hyun Yoon.
\newblock In-datacenter performance analysis of a tensor processing unit.
\newblock In {\em Proceedings of the 44th Annual International Symposium on
  Computer Architecture}. {ACM}, June 2017.

\bibitem{Lee2018}
Jinmook Lee, Changhyeon Kim, Sanghoon Kang, Dongjoo Shin, Sangyeob Kim, and
  Hoi-Jun Yoo.
\newblock Unpu: A 50.6tops/w unified deep neural network accelerator with
  1b-to-16b fully-variable weight bit-precision.
\newblock In {\em 2018 IEEE International Solid - State Circuits Conference -
  (ISSCC)}, pages 218--220, 2018.

\bibitem{Ando2018}
Kota Ando, Kodai Ueyoshi, Kentaro Orimo, Haruyoshi Yonekawa, Shimpei Sato,
  Hiroki Nakahara, Shinya Takamaeda-Yamazaki, Masayuki Ikebe, Tetsuya Asai,
  Tadahiro Kuroda, and Masato Motomura.
\newblock Brein memory: A single-chip binary/ternary reconfigurable in-memory
  deep neural network accelerator achieving 1.4 tops at 0.6 w.
\newblock {\em IEEE Journal of Solid-State Circuits}, 53(4):983--994, 2018.

\bibitem{patterson2021carbonimpact}
David Patterson et~al.
\newblock The carbon footprint of machine learning.
\newblock {\em Communications of the ACM}, 64(7):16--18, 2021.

\bibitem{10.1145/3297858.3304049}
Aayush Ankit, Izzat~El Hajj, Sai~Rahul Chalamalasetti, Geoffrey Ndu, Martin
  Foltin, R.~Stanley Williams, Paolo Faraboschi, Wen-mei~W Hwu, John~Paul
  Strachan, Kaushik Roy, and Dejan~S. Milojicic.
\newblock Puma: A programmable ultra-efficient memristor-based accelerator for
  machine learning inference.
\newblock In {\em Proceedings of the Twenty-Fourth International Conference on
  Architectural Support for Programming Languages and Operating Systems},
  ASPLOS '19, page 715–731, New York, NY, USA, 2019. Association for
  Computing Machinery.

\bibitem{Drakakis1999}
E.M. Drakakis, A.J. Payne, and C.~Toumazou.
\newblock "log-domain state-space": a systematic transistor-level approach for
  log-domain filtering.
\newblock {\em IEEE Transactions on Circuits and Systems II: Analog and Digital
  Signal Processing}, 46(3):290--305, 1999.

\bibitem{Yang2015}
Guang Yang, Richard~F. Lyon, and Emmanuel.~M. Drakakis.
\newblock A $6 \mu w$ per channel analog biomimetic cochlear implant processor
  filterbank architecture with across channels agc.
\newblock {\em IEEE Transactions on Biomedical Circuits and Systems},
  9(1):72--86, 2015.

\bibitem{Harrison1978}
David Harrison and Daniel~L Rubinfeld.
\newblock Hedonic housing prices and the demand for clean air.
\newblock {\em Journal of Environmental Economics and Management},
  5(1):81--102, March 1978.

\bibitem{scikit-learn_boston}
The scikit-learn developers.
\newblock The scikit-learn boston housing dataset documentation.
\newblock 2021.
\newblock Accessed: 2024-12-29.

\bibitem{fairlearn_boston}
Fairlearn Developers.
\newblock The boston housing dataset and fairness concerns.
\newblock \textit{Fairlearn Documentation}, 2021.
\newblock Accessed: 2024-12-29.

\bibitem{8556738}
Ali~Al Bataineh and Devinder Kaur.
\newblock A comparative study of different curve fitting algorithms in
  artificial neural network using housing dataset.
\newblock In {\em NAECON 2018 - IEEE National Aerospace and Electronics
  Conference}, pages 174--178, 2018.

\bibitem{Gerosa2004}
A.~Gerosa, A.~Maniero, and A.~Neviani.
\newblock A fully integrated dual-channel log-domain programmable preamplifier
  and filter for an implantable cardiac pacemaker.
\newblock {\em IEEE Transactions on Circuits and Systems I: Regular Papers},
  51(10):1916--1925, 2004.

\bibitem{Seevinck1990}
E.~Seevinck.
\newblock Companding current-mode integrator: A new circuit principle for
  continuous-time monolithic filters.
\newblock {\em Electronics Letters}, 26:2046--2047, 1990.

\bibitem{Moro-Fria2011}
D.~Moro-Frias, M.~T. Sanz-Pascual, and C.~A. de~la Cruz~Blas.
\newblock A novel current-mode winner-take-all topology.
\newblock In {\em 2011 20th European Conference on Circuit Theory and Design
  (ECCTD)}, pages 134--137, 2011.

\end{thebibliography}

\end{document}